\documentclass[10pt,letterpaper]{article}
\usepackage{opex3}
\usepackage{amsmath}
\usepackage{cite}

\begin{document}

\title{Excitation of atoms in an optical lattice driven by polychromatic amplitude modulation}

%
%
\author{Linxiao Niu, Dong Hu, Shengjie Jin, Xiangyu Dong, Xuzong Chen, and Xiaoji Zhou$^{*}$}

\address{School of Electronics Engineering $\&$ Computer
Science, Peking University, Beijing 100871, China}

\email{$^*$xjzhou@pku.edu.cn}


\begin{abstract}
We investigate the mutiphoton process between different Bloch states
in an amplitude modulated optical lattice. In the experiment, we
perform the modulation with more than one frequency components,
which includes a high degree of freedom and provides a flexible way
to coherently control quantum states. Based on the study of single
frequency modulation, we investigate the collaborative effect of
different frequency components in two aspects. Through double
frequency modulations, the spectrums of excitation rates for
different lattice depths are measured. Moreover, interference
between two separated excitation paths is shown, emphasizing the
influence of modulation phases when two modulation frequencies are
commensurate. Finally, we demonstrate the application of the double
frequency modulation to design a large-momentum-transfer beam
splitter. The beam splitter is easy in practice and would not
introduce phase shift between two arms.
\end{abstract}


\section{Introduction}
Ultracold atoms in periodically driven optical lattices, including
shaken and amplitude modulated systems, have attracted much
attention recent years, for they can bring out new interesting
phenomena including realization of artificial gauge fields in
different lattice geometries~\cite{G1,G2,G3} and the coherent
control of atomic wavefunctions~\cite{A,B,C}. In a shaken one
dimensional optical lattice, the ferromagnetic transition in trapped
Bose gas has also been observed~\cite{Colin} by coupling ground band
$s$ of the lattice system to the first excited band $p$. Specific to
amplitude modulated lattice systems, although only bands with the
same parity can be coupled, there are still researches on a wide
range of problems, such as transfer of atoms from the ground band
$s$ to the second excited band $d$~\cite{WDP}, detection of
superfluid-Mott insulator transition~\cite{TFA} and study of the
dynamical tunnelling of ultracold atoms with quantum
chaos~\cite{DTO,DTO2}. The technique can also be applied in areas
including realization of a velocity filter~\cite{MWS} and detection
of gravity~\cite{AWP,PMO}.

Normally, studies of amplitude modulated lattices are based on
single frequency modulations that only involve processes with one
photon emission or absorption. During a polychromatic modulation,
not only the amplitudes, but also the phases of different frequency
components in the modulation can be controlled independently,
providing a more flexible way to coherently manipulate quantum
states. In this paper, we coherently transfer atoms from the ground
state $s$ to the high excited $g$ band via a double frequency
modulation. The peaks of transfer rate with different lattice depths
are measured experimentally, while influence of the modulation phase
is demonstrated by performing an interference between two
independent paths of excitations. These experiments completely
investigated how frequencies and phases of different modulation
frequency components would influence the excitation between Bloch
states. Furthermore, we show an application of the double frequency
modulation to build a large-momentum-transfer (LMT) beam splitter.
Comparing with LMT beam splitters based on Bloch
oscillation~\cite{WDP,LMBS} or high order Bragg
scattering~\cite{LMBS2}, our method is easy in practise and would
not introduce phase shift between two separated atom clouds.

The paper is organized as follows: In Sections~\ref{S2} and \ref{S3} the
polychromatic modulation theory based on Floquet method and our
experimental system are introduced respectively. In Section~\ref{S4},
we study the effects of the modulation phase for the single
frequency modulation. In Section~\ref{S5} collaborative effect of different
frequency components is studied, demonstrate the effect
of both modulation frequencies and phases. Section~\ref{S6} presents a way
to realize a LMT beam splitter. Discussion and conclusion are in Section~\ref{S7}.

\section{Theory for optical lattice with polychromatic modulation}\label{S2}
\begin{figure}
\begin{center}
\includegraphics[bb=141 565 424 695,width=7cm]{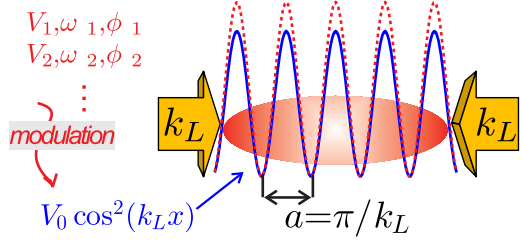}
\end{center}
\caption{A sketch of lattice depth modulation in our system. The
depth of lattice potential $V_0\cos^2(k_{L}x)$ is driven by a
polychromatic modulation
$\sum_iV_i\cos(\omega_it+\phi_i)$.}\label{fig1}
\end{figure}

For an atom in an amplitude modulated lattice system along $\hat{x}$
axis, as schematically shown in Fig.~\ref{fig1}, the time
dependent Hamiltonian can be written as
\begin{equation}\label{H}
H(t)=\frac{p^2_x}{2M}+V_{0}\cos^2(k_{L}x)+\sum_i V_{i}\cos(\omega_i
t+\phi_i)\cos^2(k_{L}x).
\end{equation}
The first term in the right hand side is kinetic energy with $M$ the
atom's mass and $p_x$ its momentum along the $x$ direction. The second
term represents optical lattice without the modulation. $V_0$ is the
constant part of the lattice depth, and the wave vector is
$k_L=2\pi/\lambda$ with $\lambda$ the laser wavelength. The last
term expresses the amplitude modulation with the modulation
amplitude $V_i$, the frequency $\omega_i$ and the phase $\phi_i$ of
each frequency components.

Typically, time-periodic systems are described by Floquet's
theorem~\cite{FBT}. In our system, each of the modulation
frequencies $\omega_i$ gives a period $T_i=2\pi/\omega_i$ and the
Hamiltonian has a period of $H(t+T)=H(t)$, with $T$ the lowest
common multiple of $T_i$. By defining a time-evolution operator for
one period as $\hat{U}(T)$, solutions to this problem must satisfy
\begin{equation}\label{H1}
|\psi_{q,\alpha}(t+T)\rangle=\hat{U}(T)|\psi_{q,\alpha}(t)\rangle=
e^{i\epsilon_{q,\alpha} T}|\psi_{q,\alpha}(t)\rangle,
\end{equation}
and the Floquet states $|u_{q,\alpha}\rangle$ are defined as
$|\psi_{q,\alpha}\rangle=e^{iqx-i\epsilon_{q,\alpha}
t}|u_{q,\alpha}\rangle$ with the quasi-momentum $q$, the band index
$\alpha$ and quasi-energy $\epsilon_{q,\alpha}$, which leads to
\begin{equation}\label{H2}
(H(t)-i\partial_t)|u_{q,\alpha}\rangle=H_{0}|u_{q,\alpha}\rangle=\epsilon_{q,\alpha}|u_{q,\alpha}\rangle.
\end{equation}

For a special case of single frequency modulation, considering the
time and coordinate periodicity of the Floquet states
$|u_{q,\alpha}\rangle$, by a Fourier transformation we can study the
problem with a set of basis {$|v_{lm}\rangle=e^{i (2lk_Lx-m\omega_1
t)}$} in an extended Hilbert space
$\mathcal{S}=\mathcal{H}\otimes\mathcal{T}$. Where $\mathcal{H}$ is
the Hilbert space and $\mathcal{T}$ is the space of all functions
with periodic $T$. By integration in one period, we can turn to a
time-independent system.
\begin{equation}\label{HM}
A_{lm,l^\prime m^\prime}=\frac{1}{T}\int_0^T|v_{l^\prime
m^\prime}\rangle H_0\langle v_{lm}|dt.
\end{equation}

To get rid of the phase $\phi_1$, we perform an unitary
transformation $U_{lm,lm}=e^{im\phi_1}$ to the Hamiltonian by
$A^{\prime}=U^{-1}A U$. There are three kinds of terms in the matrix
of $A^{\prime}$. The diagonal terms $A^{\prime}_{lm;lm}=(2l\hbar
k_L)^2/2M+m\hbar\omega_1$ are the energy of momentum states shifted by
absorbing or emitting Floquet photons with energy $\hbar \omega_1$.
The terms $A^{\prime}_{lm;l\pm1,m}=V_0/4$ show stationary component
of the lattice would coupling atoms with momentum difference $2\hbar
k_L$. In addition, $A^{\prime}_{lm;l\pm1,m\pm1}=V_1/8$ terms show the
time modulation of the lattice, which would coupling two momentum
states separated by $2\hbar k_L$ while absorbing or emitting one
Floquet photon.

When the modulation frequency is near-resonant with the energy difference
between two specific bands and far detuned from the others, we can get an
effective Hamiltonian by a rotating wave approximation.

\begin{equation}\label{sHM}
H_R=\left(
  \begin{array}{cc}
    E_\alpha & V_1e^{i\phi_1}\Omega_{\alpha\beta} \\
    V_1e^{-i\phi_1}\Omega_{\alpha\beta}^* & E_\beta-\hbar \omega \\
  \end{array}
\right),
\end{equation}
where $E_\alpha$ and $E_\beta$ are the energy of Bloch states $|\alpha\rangle,
|\beta\rangle$ in the system without modulation. Coupling constant
between two states is $\Omega_{\alpha
\beta}=\langle \alpha|\cos^2({k_Lx})|\beta\rangle/4$.

The extension to polychromatic driven is straightforward. For
example, a double frequency modulation induced two-photon process
between $s$-$g$ band is described by an effective Hamiltonian
$H_{sg}$ as:
\begin{small}
\begin{eqnarray}\label{dHM}
&&\!H_{sg}= \nonumber\\
&&\left(
\begin{array}{cccccc}
E_s & e^{i\phi_1}V_1\Omega_{sd} & e^{i\phi_2}V_2\Omega_{sd} & 0 & 0 & 0\\
\!\! e^{-i\phi_1}V_1\Omega_{sd}^* & E_d-\hbar \omega_1 & 0 & e^{i\phi_2}V_2\Omega_{dg} & e^{i\phi_1}V_1\Omega_{dg} & 0\\
\!\! e^{-i\phi_2}V_2\Omega_{sd}^* & 0 & E_d-\hbar \omega_2 & e^{i\phi_1}V_1\Omega_{dg} & 0 & e^{i\phi_2}V_2\Omega_{dg}\\
0 & e^{-i\phi_2}V_2\Omega_{dg}^* & e^{-i\phi_1}V_1\Omega_{dg}^* & E_g-\hbar(\omega_1+\omega_2) & 0 & 0\\
0 & e^{-i\phi_1}V_1\Omega_{dg}^* & 0 & 0 & E_g-2\hbar\omega_1& 0\\
0 & 0 & e^{-i\phi_2}V_2\Omega_{dg}^*& 0 & 0& E_g-2\hbar\omega_2\\
\end{array}
\!\!\!\right),
\end{eqnarray}
\end{small}

The effective Hamiltonian $H_{sg}$ is constructed by means of nearly
degenerate perturbation technique~\cite{NDP}, in which we include six
nearly degenerate states considering four main processes in the
excitation. The six states are $|E_s\rangle$ the $s$
band, $|E_d-\hbar\omega_1\rangle$, $|E_d-\hbar\omega_2\rangle$ the
$d$ band dressed by Floquet photon $\omega_1$ or $\omega_2$ and
$|E_g-\hbar(\omega_1+\omega_2)\rangle$, $|E_g-2\hbar\omega_1\rangle$
and $|E_g-2\hbar\omega_2\rangle$ the $g$ band dressed by two Floquet
photons. Using this basis a general state
$(v_1,v_2,v_3,v_4,v_5,v_6)^T$ gives complex coefficient of the six
dressed states. Population of Bloch states $s$ is $|v_1|^2$, while
population on $g$ band is $|v_4e^{i(\omega_1+\omega_2) t}+v_5
e^{2i\omega_1 t}+v_6 e^{2i\omega_2 t}|^2$, given by coherent
superposition of all $g$ band states dressed with different Floquet
photons. Solution of the model consists with the time dependent
Schr\"odinger equation and the effective model provides us a better
understanding of the mutiphoton process. However,
in the calculation more states associated with higher order
processes could be included to get a more accurate result,
especially when the modulation amplitude is large.

\section{Experimental system}\label{S3}

Our experiment begins with a quasi-pure condensate of typically
$1.5\times10^5$ $^{87}Rb$ atoms in the $|F=2,m_F=2\rangle$ hyperfine
ground state, produced in a combined potential of a single-beam
optical dipole trap and a quadruople magnetic trap. The trapping
frequencies are $\omega_x=2\pi\times 28Hz, \omega_y=2\pi\times 60Hz,
\omega_z=2\pi\times 70Hz$. The optical lattice is formed by a
retro-reflected red detuned laser beam, with lattice constant
$a=\lambda/2=426nm$ focused to a waist of $110\mu m$. Density of
atoms in our system is less than $5\times10^{13}cm^{-3}$, and the
mean-field interaction can be omitted to capture the main physical
mechanism in the excitation~\cite{intera,intera2}.

The lattice depth is calibrated by Kapitza-Dirac scattering and the
modulation of lattice depth is controlled by an acousto-optic
modulator(AOM). The modulation amplitudes $V_i$ and phases $\phi_i$
are generated from a signal generator and the intensity of lattice
laser is monitored by a photodetector. Experimental results are
absorption images taken after $28ms$ time-of-flight (TOF).
Occupation number at different momentum states $|2l\hbar k_L\rangle$
($l$ is integral momentum index) can be given from TOF images by
$n_l=N_l/N$, with $N_{l}$ the atom number at momentum state
$|2l\hbar k_L\rangle$ and $N$ the total atom number. The initial
state for experiment is prepared non-adiabatically with numerically
designed sequence of lattice pulses~\cite{Liu,Zhai}.  For
experimental convenience, the lattice pulses are carried out with
the same depth as the constant part of the modulated lattice
potential $V_0$. Typically, each of the pulses and the subsequent intervals
are lasting for no more than $25\mu s$, and the whole process can be
finished within $60\mu s$, thus the loading time is greatly reduced
comparing with traditional adiabatic loading method.

\section{The initial phase effect in single frequency modulation}\label{S4}
Single frequency modulation can be seen as the basis of
polychromatic driven. In this part, we present the preparation of a
Floquet state in the single frequency driven system, which is shown
to be highly related to the modulation phase.

\begin{figure}
\begin{center}
\includegraphics[bb=111 321 445 488,width=8cm]{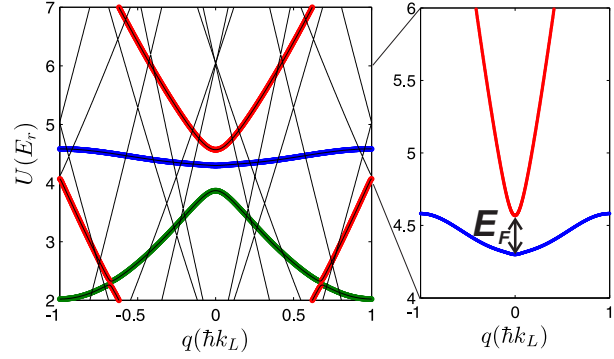}
\end{center}
\caption{Left side is the calculated Floquet spectra of a single
frequency driven system, with parameters $V_0=5E_r$, $V_1=0.5E_r$,
$\hbar\omega_1=5E_r$. In the figure the first seven bands are
presented. The heavy lines depict states maximally overlapping with
the $s$(blue), $p$ (green) and $d$ (red) Bloch bands respectively.
Right side shows the details of two Floquet bands most overlapping
with $s$ and $d$ bands. The two bands are separated by a band gap
$E_F$ at $q=0$.}\label{fig2}
\end{figure}

Following the discussion in Sec.~\ref{S2}, Fig.~\ref{fig2} depicts a
typical quasi-energy spectrum of the single frequency driven system,
which is obtained by direct diagonalization of $A^{\prime}$ at
various quasi-momentum $q$. The same calculation also gives
eigenvectors in the extended Hilbert space. The spectrum exhibits a
complex structure as a result of the periodically repetition of high
excited bands.

To connect our calculation with the experiments, we should project
the Floquet states $|u_{q,\alpha}\rangle$ into momentum space. The
occupation number at momentum states $|2l\hbar k_L\rangle$ and its
phase can be given by a summation of all the Fourier components with
the same $l$ as $c_l(t)|2l\hbar k_L\rangle=\Sigma_m e^{-i
m\omega_1\tau}\nu_{lm}|2l\hbar k_L\rangle$, where  $\nu_{lm}=\langle
v_{lm}|u_{q,\alpha}\rangle$ is coefficient of the Floquet state.
$\tau$ is related to the modulation phase $\phi_1$ and holding time
$t$ as $\tau=\frac{\omega_1t+\phi_1}{2\pi}T$. It is also useful to
define the overlapping between the Floquet state
$|u_{q,\alpha}\rangle$ and a Bloch state $|n_q\rangle$ of the
undriven potential as $P=\int_0^T|\sum_l\langle c_l(t)|n_q\rangle|^2
dt$. Property of the Floquet band is typically characterized by its
most overlapping Bloch band~\cite{TTM}. Without loss of generality,
our experiments are restricted to quasi-momentum $q=0$, and energy
gap between Bloch bands $\alpha$ and $\beta$ are written as
$\hbar\omega_{\alpha\beta}$. The technique can also be applied in
systems with acceleration~\cite{MWS}, which brings out phenomena different from
our study.

When modulation phase $\phi_1$ is given, a Floquet state can be
projected into the momentum space, and such a state can be prepared
by carrying out two lattice pulses with numerically designed pulse
sequence~\cite{Liu,Zhai}. In the fast loading process, lattice depth
of the two pulses is kept constant. For a target state, the fidelity
of a prepared state can be given numerically for different pulse
durations and time intervals, and the optimized pulse sequence is
obtained by finding the maximum loading fidelity. Throughout this
method we can get a loading fidelity of more than $95\%$
experimentally.

\begin{figure}
\begin{center}
\includegraphics[bb=80 205 449 676,width=7cm]{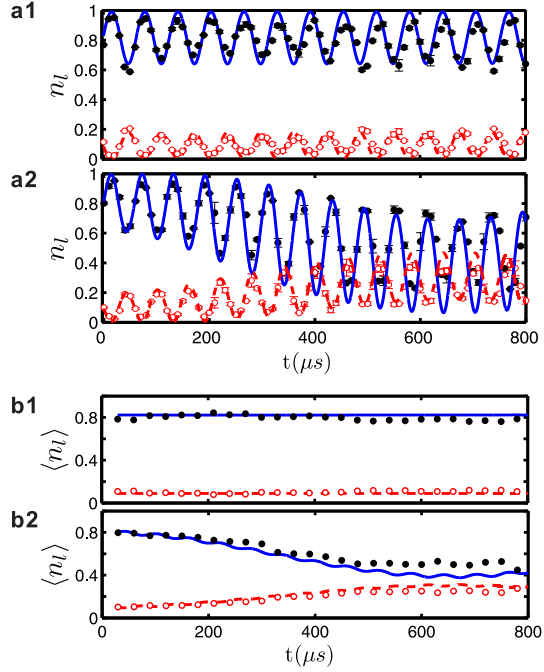}
\end{center}
\caption{Time evolution of $n_l$ measured from the experiments with
initial modulation phase (a1) $\phi=-\pi/2$ and (a2) $\phi=\pi/2$.
Time averaged fraction $\langle n_l\rangle$ are also shown for (b1)
$\phi=-\pi/2$ and (b2) $\phi=\pi/2$ respectively. $n_0$ is shown
with black dots comparing to the numerical simulation in solid
lines. $n_{1}$ and $n_{-1}$ are shown in average with red circles
the corresponding numerical result is shown in dashed lines. Each
point is averaged by three experiments and the error bars indicate
the standard deviation.  }\label{fig3}
\end{figure}

With the initial modulation phase $\phi_1=-\pi/2$, a pure Floquet
state is loaded by lattice pulses calculated with
$\tau=\frac{\phi_1}{2\pi}T=-T/4$. For the same state, we also
present a modulation with $\phi_1=\pi/2$ to show the influence of
modulation phase. The population on momentum components $|0\hbar
k_L\rangle$ and $|\pm2\hbar k_L\rangle$ for different initial phases
$\phi_1=-\pi/2$ and $\phi_1=-\pi/2$ are shown in
Fig.~\ref{fig3}(a1) and \ref{fig3}(a2) respectively. The parameters are $V_0=5.0E_r,
V_1=0.5E_r$ and driven frequency is $\hbar\omega_1=5.0E_r$, red
detuned from the band gap $\hbar\omega_{sd}=5.2E_r$.
Figure~\ref{fig3}(a1) is a Floquet state, which shows a time period of
$T=2\pi/\omega_1=62.75\mu s$. The time evolution in momentum space
would be quite different when the modulation phase is changing by
$\pi$, as shown in Fig.~\ref{fig3}(a2). Besides the oscillation with
a frequency near $\omega_1$, we observe a slow growth of $n_1$.
Meanwhile, the amplitude of oscillations are also increasing
following increase of the population on $d$ band.

The influence of modulation phase is shown more clearly when taking
time average in one modulation period $T$. In Fig.~\ref{fig3}(b1)
with $\phi_1=-\pi/2$ the Floquet state shows a constant population
in different momentum states by taking time average. While for
$\phi_1=\pi/2$, Fig.~\ref{fig3}(b2) shows a Rabi oscillation between
$s$-$d$ bands with Rabi frequency $E_F/\hbar$, where $E_F$ is the
gap between two Floquet bands. The observation can also be explained
by Eq.~(\ref{sHM}). The phase of coupling terms would change with
$\phi_1$, thus the eigenstate is superposition of $|s\rangle$ and
$|d\rangle$ with a relative phase determined by $\phi_1$. When the
modulation phase is changed, the loaded state is no longer an
eignstate, and we can observe the Rabi oscillation.

\section{Excitation of ground state via a double frequency modulation}\label{S5}
The study of lattice modulation can be extended from single
frequency to a more general form as described in Eq.~(\ref{dHM}).
Specific to two-photon excitation between $s$-$g$ band, the time
dependent lattice is described as $V_L(t)=V_0+V_1\cos(\omega_1
t+\phi_1)+V_2\cos(\omega_2 t+\phi_2)$, which includes seven
parameters $V_0,V_1,V_2,\omega_1,\omega_2,\phi_1,\phi_2$. During the
experiments, we mainly focus on the influence of modulation
frequencies and phases, while keeping other parameters constant.

\begin{figure}
\begin{center}
\includegraphics[bb=35 555 475 698,width=9cm]{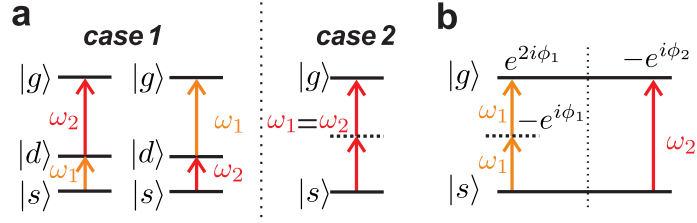}
\end{center}
\caption{(a) Two special cases in detecting the transfer population
spectrum. In case 1, absorption of photons with $\omega_1$(orange)
or $\omega_2$(red) is resonant with $d$ band. In case 2, two
frequencies are equal. (b) For $s$-$g$ coupling $\omega_1$ provides a
two-photon process while $\omega_2=2\omega_1$ provides a one-photon
process. Phases of two paths are controlled independently by
modulation phases of $\omega_1$ and $\omega_2$. }\label{fig4}
\end{figure}

\subsection{Spectrum of two-photon excitation}

The single frequency modulation shows that $\phi_1$ is related to
relative phase between $s$ and $d$ band components of the Floquet
state. However, with a nearly pure $s$ band, the phase of single
modulation can be neglected, only the phase difference between two
modulations is important. Therefore, we chose $\phi_1=\pi$ while
leaving $\phi_2$ variable to control the relative phase between two
modulations.

For $s$-$g$ band coupling through a two-photon absorption process,
the sum of the two modulation frequencies is chosen as
$\omega_1+\omega_2=\omega_{sg}$. This process is well described by
Eq.~(\ref{dHM}), in which we have considered different pumping paths,
as schematically shown in Fig.~\ref{fig4}(a). There are two cases
which would benefit the excitation process.

{\bf Case 1:} Resonant two-photon process. When
$\omega_1=\omega_{sd}$ or $\omega_2=\omega_{sd}$, atoms are
transferred from $|s\rangle$ to $|g\rangle$ with the assistance of
$d$ band as an intermediate band.

{\bf Case 2:} Equal frequency two-photon process. When
$\omega_1=\omega_2=\omega_{sg}/2$, two modulations with the same
frequency can be added together, and the coupling strength of the
process is doubled.

\begin{figure}
\begin{center}
\includegraphics[bb=128 195 411 587,width=7cm]{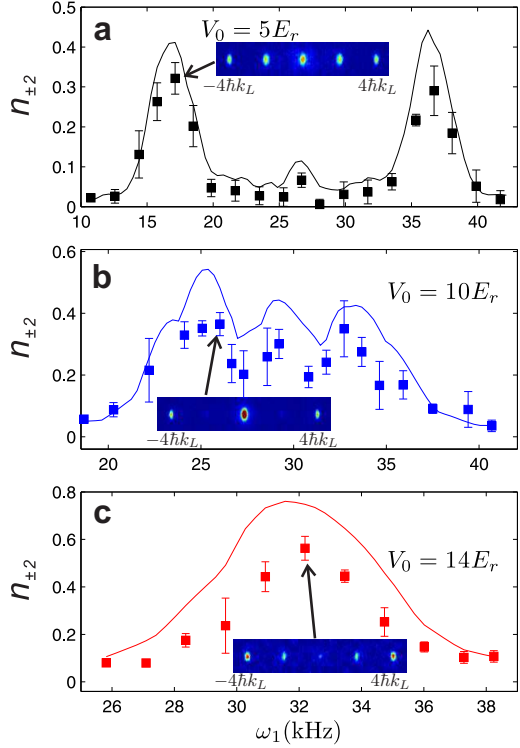}
\end{center}
\caption{Spectrum for the population on $\pm4\hbar k_L$ states with
increasing of modulation frequency $\omega_1$. Population detected
on $\pm4\hbar k_L$ after a double frequency modulation for (a)
$V_0=5E_r$(black) with $V_1=1.4E_r$, $V_2=1.6E_r$ $t=300\mu s$, (b)
$V_0=10E_r$(blue) with $V_1=2.8E_r$, $V_2=2.2E_r$ and $t=200\mu s$,
(c) $V_0=14E_r$(red) with $V_1=V_2=2.5E_r$ and $t=150\mu s$ are
shown in rectangles with error bars. Solid lines are corresponding
numerical simulation.}\label{fig5}
\end{figure}

With resonance condition $\omega_1=\omega_{sd}$ in {\bf case 1},
when two modulation amplitudes are chosen as
$V_1\Omega_{sd}=V_2\Omega_{dg}$ the transfer rate would show a
maximum resonant peak. And we keep this modulation amplitudes while
sweeping $\omega_1$. Modulation phase is chosen from numerical
simulation to get a maximum transfer.

In the experiments, we sweep the frequency $\omega_1$ for different
lattice depths $V_0=5E_r$, $10E_r$ and $14 E_r$. Population on
momentum states $\pm4\hbar k_L$ measured from the experiment are
shown with rectangles in Fig.~\ref{fig5}, comparing with the
theoretical calculation shown in solid curves. Within the lattice
depth we considered, $g$ band is greatly concentrated on $|\pm4\hbar
k_L\rangle$ momentum states, thus $n_{\pm2}$ can reflect transfer
rate to $g$ band.

Figure~\ref{fig5}(a) shows the case of $V_0=5E_r$. For the lattice
depth we have $\Omega_{sd}/\Omega_{dg}=1.11$, correspondingly the
modulation amplitudes are chosen as $V_1=1.4E_r$, $V_2=1.6E_r$, and
$t=300\mu s$. The holding time $t$ may be chosen shorter if the
maximum of numerical simulation is reached at an earlier time. In the
figure, two peaks appear at $\omega_1=\omega_{sd}$ and
$\omega_2=\omega_{sd}$ which follows {\bf case 1} we have discussed.
And there the central peak at frequency $\omega_1=\omega_2$
following {\bf case 2} is much lower than two peaks for {\bf case 1}.

Figure~\ref{fig5}(b) shows the spectrum with $V_0=10E_r$.
With the increasing of $V_0$, the
energy difference $\omega_{sd}$ is getting closer to $\omega_{dg}$, and
three peaks are overlapping. Comparing with $V_0=5E_r$ the central
peak is much higher, because the process of {\bf case 2} is also
near resonance with $d$ band.

For $V_0=14E_r$ only one peak would be measured in the spectrum as
shown in Fig.~\ref{fig5}(c), which means $\bf case 1$ and $\bf case 2$
are fulfilled simultaneously. Under this condition, the coupling
between $s$ and $g$ band is also greatly enhanced. Modulation
amplitudes are $V_1=V_2=2.5E_r$, for the two modulation frequencies are
the same at $\bf case 1$ and can't be distinguished.

The experimentally detected peaks are governed by two cases, which
are within the description of Eq.~(\ref{dHM}). Thus in the numerical
simulation we can neglect higher order processes of emission and
absorption of Floquet photons. The discrepancy between experimental
result and the theoretical simulation is probably due to the
influence of interaction and initial momentum distribution of the
condensate. These effects would destroy coherency during the
modulation, and the measured population of excited state would be
lower than the maximum value in the theoretical simulation.

\subsection{The role of modulation phases}

\begin{figure}
\begin{center}
\includegraphics[bb=15 520 531 700,width=12cm]{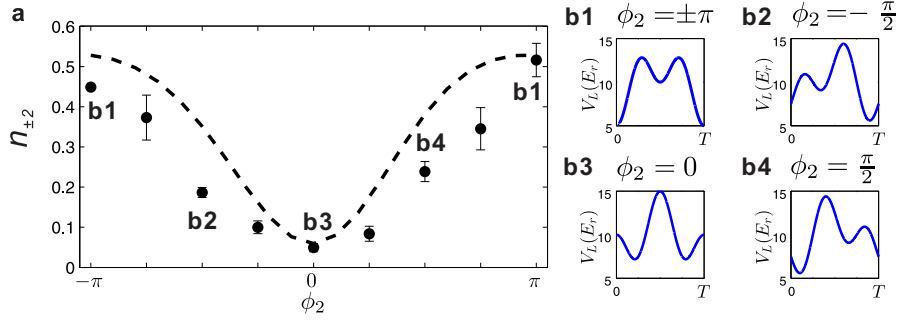}
\end{center}
\caption{The excited population on $g$ band shows the interference
between two paths. (a) Population transferred to $n_{\pm2}$ is shown in
black dots with error bars. The dashed line shows theoretical
simulation for comparison. (b1)-(b4) $V_L$ for different
phases.}\label{fig6}
\end{figure}

When $\omega_1$ and $\omega_2$ are incommensurate, the influence of
relative phase is not prominent because there is only one path for
the pumping process and no states could interfere with each other.
However, the effect of modulation phases would be more pronounced
when two modulation frequencies are commensurate.

As shown in Fig.~\ref{fig4}(b), when $2\omega_1=\omega_2$, relative
phase of  modulations can be shown by performing a one-photon
process simultaneously with a two-photon process. We choose the
frequency $\omega_1=\omega_{sg}/2$ at central peak of $V_0=10E_r$
lattice and the second modulation is performed with
$\omega_2=\omega_{sg}$. Modulation amplitudes are $V_1=V_2=2.5E_r$
and $t=500\mu s$. Similar to the process described by Eq.~(\ref{dHM}),
in this problem we consider the interference between two states
$|E_g-2\hbar\omega_1\rangle$ and $|E_g-\hbar\omega_2\rangle$. During
the two-photon process with $\phi_1=\pi$, phase of state
$|E_g-2\hbar\omega_1\rangle$ remains zero, while $\phi_2$ determine
phase of state $|E_g-\hbar\omega_2\rangle$ as $\phi_2-\pi$. Relative
phase of two modulations is defined as
$(\phi_2-\pi)-\frac{\omega_2}{\omega_1}(\phi_1-\pi)=\phi_2-\pi$, and
the interference can be changed from constructive to destructive
with different $\phi_2$.

Figure~\ref{fig6}(a) depicts the population transferred to $g$ band
with the phase of one-photon process changing by $2\pi$.
Figure~\ref{fig6}(b) shows how the depth of lattice is varying
with time for four different modulation phases. With $\phi_2=\pi$,
the population transferred to $|g\rangle$ through one-photon process
and through two-photon process are in phase and the transfer rate is
enhanced. When $\phi_2$ is increasing, relative phase of two
processes are deviating from zero, and the population on $|g\rangle$
would decrease. In the case of $\phi_2=0$ the two process are out of
phase, and only few atoms can be transferred to $g$ band. Further
increasing of $\phi_2$ would increase the measured population, and
when $\phi_2$ is changed by $2\pi$ the population on $g$ band
reaches the maximum value again.

The single atom Hamiltonian in Eq.~(\ref{H}) can well explain the
excitation in our system. However, for the $500\mu s$ modulation,
decoherence from mean-field interaction and the momentum
distribution becomes significant, thus it is necessary to carry out
a simulation based on the time-dependent Gross-Pitaevskii equation,
\begin{eqnarray}\label{GPE}
i\hbar\dfrac{{\partial \psi}}
{{\partial t}} = [-\frac{\hbar^{2}}{2 m}\frac{{\partial^2}}{{
\partial {x^2}}}+ V_{L}(x,t)+\frac{1}{2}m\omega_x^{2}
x^{2}+g{\left|\psi\right|}^{2}]\psi,
\end{eqnarray}
where $g$ is the parameter of interaction. In the simulation, we
consider both the distribution of initial momentum and the
mean-field interaction which would deduce the measured
excitation rate. Dashed line in Fig.\ref{fig6}(a) gives the result
of simulation which fits well with the experiment. Similarly, the
relative phase can also be changed by $\phi_1$ of the two-photon
process, which gives a period of $\pi$.

The collaborative effect of different frequency components is studied in
two aspects, demonstrate the effect of both modulation frequencies and
phases of different frequency components. The usefulness of this study
is not limited to double frequency modulation, within these two aspects,
collaborate effect of more frequency components can also be well understood.

\begin{figure}
\begin{center}
\includegraphics[bb=141 398 471 579,width=7cm]{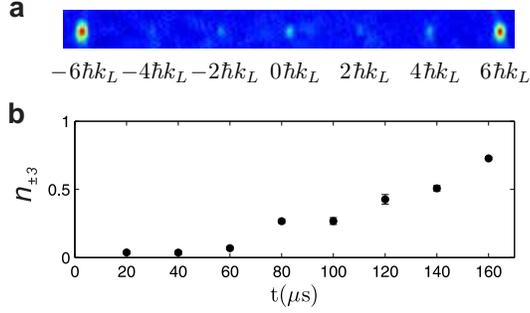}
\end{center}
\caption{A LMT beam splitter with a separation of $12\hbar k_L$. (a)
TOF image of the LMT beam splitter. (b) Experimentally measured
population of atom on momentum states $|\pm6\hbar k_L\rangle$ are
shown in black dots with error bars.}\label{BS}
\end{figure}

\section{Application in LMT beam splitter}\label{S6}

The double frequency modulation can also be applied to realize
a LMT beam splitter by pumping atoms to higher momentum states,
which is useful in experiments of atomic interferometry
~\cite{interferometer}. According to the resonant condition we
have discussed, a preferred choice is $V_0=14E_r$ with the
modulation frequency $\omega_1=\omega_{sd}=\omega_{dg}$. The
other modulation frequency is chosen as $\omega_2=\omega_{gi}$
(with $i$ the $6^{th}$ excited Bloch band) to get a distribution
at $\pm6\hbar k_L$.

We begin with a condensate, the lattice is suddenly turn on,
modulated with $V_1=V_2=2.5E_r$ and the preferable phases are found
numerically. A TOF image of experimental result is shown in
Fig.~\ref{BS}(a). Figure~\ref{BS}(b) shows the population of atoms on
momentum states $|\pm6\hbar k_L\rangle$ where we have subtracted the
thermal gas. Near $80\%$ of the atoms are coherently transferred
into $\pm 6\hbar k_L$ momentum states within $160\mu s$, and the
maximum lattice depth needed is below $V_0+V_1+V_2=24E_r$. Comparing
with a LMT beam splitter based on high order Bragg
scattering~\cite{LMBS2}, our method needs a much lower lattice depth
and is easy in practice. Furthermore, the process is symmetric for both sides,
and would not introduce phase shift between two separated atom
clouds.

A momentum splitting of $12\hbar k_L$ is not the limit of the amplitude
modulation. More frequency components or another subsequent
modulation can be introduced to reach a larger momentum splitting. For
example, after the double frequency modulation, by preforming
another single frequency modulation resonance with the energy
difference between $|\pm6\hbar k_L\rangle$ and $|\pm8\hbar
k_L\rangle$ momentum states, the atoms can be transferred to
$|\pm8\hbar k_L\rangle$ coherently.

\section{Discussion and conclusion}\label{S7}

Following Floquet-Bloch theory we have presented an experimental
preparation of a Floquet state and study its property. The
non-adiabatic loading method would take much less time and greatly
reduce the heating problem. The loading method could also be
extended to systems with a shaken lattice~\cite{Colin} or a combined
modulation~\cite{C,TTM}, which provides a several millisecond longer
lifetime for condensate in the experimental study on areas including
the detection of Floquet topological states~\cite{FBT,FTS,STP}.

In conclusion, based on the study of single frequency modulation, we
investigate double frequency modulation in detail. Experimental
observations show that different modulation frequency components would
influence each other in two ways. When two frequencies are resonant with
subsequent excitation processes, the modulation can induce a resonant
two photon process which can effectively transfer atoms to higher excited
bands. With specific modulation frequencies, interference between a
one-photon path and a two-photon path is observed, revealing influence
of modulation phases when two frequencies are commensurate. The quantum
interference can be used to enhance the excitation rate or destruct
unwanted excitations. Using the
technique of double frequency modulation, we also demonstrate an
efficient way to realize a LMT beam splitter with low lattice depth.
Our study provides a more flexible way to coherently control quantum
states through an optical lattice.

\section*{Acknowledgments}
This work is partially supported by the state Key Development
Program for Basic Research of China No.2011CB921501, NSFC (Grants
No.61475007, No.11334001 and No.91336103), RFDP (Grants No.20120001110091).
\end{document}